\documentstyle[12pt,epsf]{article}
\newlength{\myleftmargin}
\newcommand{\an}{ans{\" a}tze}

\begin{document}
\thispagestyle{empty}
\begin{flushright}
{\tt DCE-APU-97-02 \\
DPNU-97-54 \\
EHU-97-11 \\
November 1997}
\end{flushright}

\vspace{0.5 cm}

\begin{center}
{\large {\bf NNI Quark-Lepton Mass Matrices in SUSY SU(5) GUT}}
\end{center}

\vspace{1cm}

\begin{center}
{\sc Toshiaki~~Ito}\footnote[2]{toshiaki@eken.phys.nagoya-u.ac.jp}~,~ 
{\sc Naotoshi~~Okamura}$^*$\footnote[3]{okamura@eken.phys.nagoya-u.ac.jp} ~and~
{\sc Morimitsu~~Tanimoto}$^{**}$\footnote[4]{tanimoto@edserv.ed.ehime-u.ac.jp}
\end{center}

\vspace{0.5 cm}

\begin{center}
\sl{Department of Childhood Education, Aichi Prefectural University,}
{\it Nagoya 467, Japan} \\
$^*$\sl{Department of Physics, Nagoya University,} {\it Nagoya 464-01, Japan} \\
$^{**}$\sl{Science Education Laboratory, Ehime University,} 
{\it Matsuyama 790, Japan}
\end{center}

\vspace{2em}

\begin{abstract}
We propose the Fritzsch$-$Branco$-$Silva-Marcos type
fermion mass matrix,
which is a typical texture
in the nearest-neighbor interaction form, in $SU(5)$ GUT.
By evolution of the mass matrices with $SU(5)$ GUT relations
 in the  minimal SUSY standard model,
we obtain predictions for the unitarity triangle of $CP$ violation as well
as the 
quark flavor mixing angles,
which are consistent with experimental data, in the case of $\tan\beta
\simeq 3$.
\end{abstract}

\clearpage

One of the most important unsolved problem of flavor physics
is the understanding of flavor mixing and fermion masses,
which are free parameters in the standard model.
The observed values of those mixing and masses may provide us 
clues to solve this problem.
Many works were made to find {\an} for quark-lepton mass matrices.
The typical one is the Fritzsch {\an} \cite{FRI},
which is called texture zero analysis
where some elements of mass matrices are required to be zero
to reduce the degrees of freedom in mass matrices.
As presented by Branco, Lavoura and Mota, both up- and down-quark mass
matrices could always be transformed to the non-Hermitian matrices
in the nearest-neighbor interaction (NNI) basis by a weak-basis transformation
for the three and four generation cases \cite{NNI}.
Based on the NNI form, several authors
have studied the quark masses and
Cabibbo-Kobayashi-Maskawa (CKM)
matrix \cite{CKM} phenomenologically \cite{BS}--\cite{TI}.
One of authors(T.I) proposed a texture, in which the up-quark mass
matrix to be
in the Fritzsch form and the down-quark mass matrix to be  
Branco$-$Silva-Marcos (BS) form \cite{TI}.
In this letter, we call this texture F-BS one.
Recently, Takasugi has shown that quark mass matrices can be transformed
in general to
either one of the following two form, the Fritzsch type parameterization or
the BS type parameterization with retaining
the NNI form for the other matrix \cite{TAKA}.
Moreover, Takasugi and Yushimura pointed out that it is reasonable
to take the BS {\an} for the down-quark mass matrix
if the up-quark one is assumed to be the Fritzsch texture.

In this letter, we build  a mass matrix model based on the F-BS texture in
$SU(5)$ GUT. 
The F-BS texture may be compared with  Georgi-Jarlskog texture \cite{GJ},
in which the zeros are forced by discrete symmetries.
We do not need to force zeros  by discrete symmetries because the  F-BS texture 
 is a typical case of the NNI form.
 
By putting $SU(5)$ GUT relations, the charged lepton mass matrix is related
with the quark one.
The evolution based on the SUSY renormalization group equations from the
GUT scale 
to the  $M_Z$ scale gives  predictions for CKM matrix.
The F-BS texture reproduces well known empirical relations,
\begin{eqnarray}
\vert V_{us}\vert &\sim& \sqrt{\frac{m_d}{m_s}} \\
\vert V_{cb}\vert &\sim& \frac{m_s}{m_b} \\
\frac{\vert V_{ub}\vert }{\vert V_{cb}\vert } &\sim& \sqrt{\frac{m_u}{m_c}} \ .
\end{eqnarray}
\noindent
Following these investigations, we take the F-BS type Yukawa
matrices  for quarks and leptons at the $SU(5)$ GUT scale.
Yukawa matrices   are written  as follows:
\begin{equation}
Y^U =
\left(
\begin{array}{ccc}
0 & a_u & 0 \\
a_u & 0 & b_u \\
0 & b_u & c_u
\end{array}
\right),
~~~
Y^D =
\left(
\begin{array}{ccc}
0 & a_de^{i\theta_1} & 0 \\
a_de^{-i\theta_1} & 0 & b_de^{i\theta_2} \\
0 & c_d & c_d
\end{array}
\right),
\end{equation}
\noindent 
 where $Y^U$ and $Y^D$ are matrices for up-quark and down-quark, respectively.
  The up-quark Yukawa matrix is the Fritzsch texture, while the
down-quark one is the BS texture.  
By assuming the ${\bf 5}^*$ and  ${\bf 45}^*$ Higgs fields
of $SU(5)$, the charged lepton Yukawa matrix $Y^E$ is given as follows:
\begin{equation}
Y^E =
\left(
\begin{array}{ccc}
0 & a_de^{i\theta_1} & 0 \\
a_de^{-i\theta_1} & 0 & -3b_d e^{i\theta_2} \\
0 & c_d & c_d
\end{array}
\right).
\end{equation}
\noindent 
Each entry of quark-lepton matrices is assumed to arise from the VEV of
${\bf 5}^*$'s
of Higgs field except for (2,3) entries
in $Y^D$ and $Y^E$, which are assumed to arise from ${\bf 45}^*$ of Higgs field.
Therefore, the matrix $Y^U$ should be symmetric while  $Y^D$ and $Y^E$ 
are allowed to be non-symmetric.
So, parameters $a_{u(d)}$, $b_{u(d)}$, $c_{u(d)}$ are taken to be  real  and
the phase parameters  appear in  $Y^D$ and $Y^E$ with $\theta_1$, $\theta_2$.
Including $\tan\beta$, we have 9 parameters in the fermion mass matrix.
On the other hand, there are 14 low energy observables, 9 charged fermion
masses,
4 CKM mixing angles and $\tan\beta$. Thus, there are 5 predictions.

We adopt a weak hypothesis for the parameters of the matrices, 
the generation hierarchy,
\begin{equation}
a_d \ll b_d \ll c_d \ .
\end{equation}
\noindent
Then, mass eigenvalues  are given in terms of  those parameters  as follows:
\begin{eqnarray}
m_d &=& \frac{a_d^2}{b_d}~~,~m_s = \frac{b_d}{\sqrt{2}}~~,~
m_b = \sqrt{2}c_d \\
m_e &=& \frac{1}{3}\frac{a_d^2}{b_d}~~,~m_\mu = 3\frac{b_d}{\sqrt{2}}~~,~
m_\tau = \sqrt{2}c_d~ .
\end{eqnarray}
\noindent
Eliminating parameters,
we obtain the $SU(5)$ GUT mass relations:
\begin{eqnarray}
m_{\tau} &=& m_b \ , \\
m_{\mu} &=& 3m_s\ , \\
m_e &=& \frac{1}{3}m_d \ ,
\end{eqnarray}
\noindent
which is the same one in Georgi-Jarlskog texture.

In the minimal SUSY standard model, the renormalization group
equations of 1-loop are \cite{RGE}
\begin{eqnarray}
\frac{d}{dt}Y^U &=& \frac{1}{16\pi^2}\Bigg\{ {\rm tr}\left(
3Y^UY^{U\dagger}\right)Y^U+3Y^UY^{U\dagger}Y^U \nonumber \\
& & \mbox{} +Y^DY^{D\dagger}Y^U-\left(\frac{13}{9}g'^2+3g_2^2
+\frac{16}{3}g_3^2\right)Y^U \Bigg\} \\
\frac{d}{dt}Y^D &=& \frac{1}{16\pi^2}\Bigg\{ {\rm tr}\left( 
Y^EY^{E\dagger}+3Y^DY^{D\dagger}\right) Y^D+3Y^DY^{D\dagger}Y^D \nonumber \\
& & \mbox{} +Y^UY^{U\dagger}Y^D-\left( \frac{7}{9}g'^2+3g_2^2+
\frac{16}{3}g_3^2\right) Y^D\Bigg\} \\
\frac{d}{dt}Y^E &=& \frac{1}{16\pi^2}\Bigg\{
{\rm tr}\left(Y^EY^{E\dagger}+3Y^DY^{D\dagger}\right) Y^E \nonumber \\
& & \mbox{} +3Y^EY^{E\dagger}Y^E-3\left( g'^2+g_2^2\right) Y^E \Bigg\}
\end{eqnarray}
for Yukawa matrices with $t=\ln{\frac{M_G^2}{\mu^2}}$ and
\begin{equation}
\frac{d}{dt}\alpha_i = \frac{b_i}{2\pi}\alpha_i^2~~~
\left( \alpha_i=\frac{g_i^2}{4\pi}~,~
g_1^2=\frac{5}{3}g'^2~,~i=1,2,3 \right),
\end{equation}
for gauge couplings, where
\begin{eqnarray}
b_1 &=& \frac{3}{10}n^{}_H+2n^{}_G, \\
b_2 &=& \frac{1}{2}n^{}_H+2n^{}_G-6, \\
b_3 &=& 2n^{}_G-9.
\end{eqnarray}

In our analysis,  the GUT scale is fixed as  $M_G=1.7\times 10^{16}$GeV
by use of the experimental data of $\alpha_1$ and $\alpha_2$.
Then we obtain $\alpha_s(M_Z)=0.114$,
which is almost consistent with the experimental data,
$\alpha_s(M_Z)=0.118\pm 0.003$ \cite{PDG}.
The factor $n^{}_H$ and $n^{}_G$ are the number of Higgs doublets
and fermion generations, respectively. We set
$n^{}_H=2$ and $n^{}_G=3$.
By numerical analysis of  the renormalization group equations,
the fermion mass matrices are obtained
at the $M_Z$ energy scale.

It is useful to comment on $\tan\beta$.
If $\tan\beta$ is less than 2, the Yukawa coupling of the $t$-quark
blows up under the GUT scale.
Recent study of the proton decay suggests that $\tan\beta$ is less than 4
\cite{PRO}.
Thus,  $\tan\beta \simeq 3$ is a reasonable region.
Actually, our numerical results favor $\tan\beta=3$.
The fits with the experimental values become worse as  $\tan\beta$ increases.

Since the two phases $\theta_1$and  $\theta_2$ in $Y^D$ and $Y^E$ hardly
affect the running
of Yukawa matrices,
the matrix elements $a_{u(d)}$,
$b_{u(d)}$, $c_{u(d)}$ can be adjusted by  the following central values of 
six fermion masses at the $M_Z$ energy scale \cite{PDG}\cite{STE},
\begin{eqnarray*}
m_u(M_Z) &=& 0.0022\pm 0.0007{\rm GeV}~,
~m_c(M_Z) = 0.59\pm 0.07{\rm GeV}~, \\
m_t(M_Z) &=& 175\pm 14{\rm GeV}~, \\
m_e(M_Z) &=& 0.486660328\pm 0.000000143{\rm MeV}~, \\
m_{\mu}(M_Z) &=& 102.7288759\pm 0.0000332{\rm MeV}~, \\
m_{\tau}(M_Z) &=& 1746.5^{+0.296}_{-0.266}{\rm MeV}~.
\end{eqnarray*}
Then  the down-quark masses are obtained as output:
\begin{displaymath}
m_d(M_Z)=0.0032{\rm GeV}~,~~m_s(M_Z)=0.081{\rm GeV}~,
~~m_b(M_Z)=3.31{\rm GeV}~ ,
\end{displaymath}
which are compared with the experimental values \cite{STE},
\begin{eqnarray*}
m_d(M_Z) &=& 0.0038\pm 0.0007{\rm GeV}~~,
~m_s(M_Z)=0.077\pm 0.011{\rm GeV}~~, \\
m_b(M_Z) &=& 3.02\pm 0.19{\rm GeV}.
\end{eqnarray*}
Thus obtained  values of down-quark masses
are almost consistent with the experimental data.
It is remarked that due to the running of Yukawa matrices,
the  (2,2) entries of the quark mass matrices
 develop in non-negligible finite ones, which are comparable
with  magnitudes of (1,2) and (2,1) entries.
On the other hand, the (1,1), (1,3) and (3,1) entries are almost zero
at the $M_Z$ scale.
So even if Yukawa matrices of quarks are F-BS type at the GUT scale,
they become to the non-NNI form at low energy scale due to the
renormalization effects.
On the other hand, the lepton mass matrix keeps the NNI form
in the renormalization running.

If $\theta_1$ and $\theta_2$ are fixed, we can predict CKM matrix.
We obtain the CKM matrix in the case of $\theta_1=253^\circ$ and
$\theta_2=60^\circ$:
\begin{equation}
\vert V_{CKM}\vert
=
\left(
\begin{array}{ccc}
0.9757 & 0.2190 & 0.0045 \\
0.2189 & 0.9747 & 0.0458 \\
0.0082 & 0.0453 & 0.9989
\end{array}
\right).
\end{equation}
This result is consistent with present experimental one within error bars .  
In particular,  we get $V_{cb}\simeq 0.045$, which should be compared with 
the prediction  $V_{cb}\geq 0.05$ in Georgi and Jarlskog texture \cite{PGJ}.
The $CP$ violating phase of the CKM matrix is also fixed.
The unitarity triangle of the predicted CKM matrix is shown in Fig. 1.
The vertex of this unitarity triangle is on the point $(0.263,0.352)$
in the $\rho$-$\eta$ plane\cite{WOL}.
Thus the vertex is in the first quadrant
in the $\rho$-$\eta$ plane as well as the case of the ref.\cite{TI}.
Here in order to describe the experimental allowed region,
we have used the following parameters \cite{Buras} and experimental data
\cite{PDG},
\begin{displaymath}
{B}_K=0.75\pm 0.15~,~~~~f_{B_d}\sqrt{B_{B_d}}=0.20\pm 0.04 \ ,
\end{displaymath}
\begin{displaymath}
\frac{\vert V_{ub}\vert}{\vert V_{cb}\vert} = 0.08\pm 0.02 \ .
\end{displaymath}
\noindent
It is noted that the predicted CKM matrix cannot reproduce
the experimental data of $\vert V_{ub}\vert / \vert V_{cb}\vert $
in the case of  $\tan\beta\geq 4$.

\begin{figure}[ht]
 \begin{center}
  \epsfbox{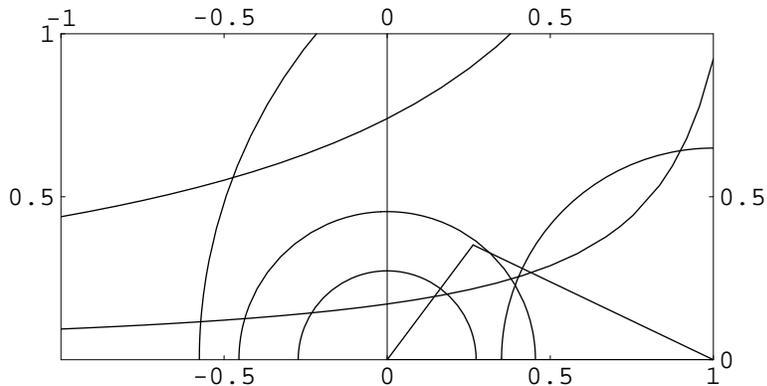}
  \caption{The unitarity triangle of CKM matrix 
in the case of $\tan\beta = 3$.  The allowed region is given by the
experimental constraints
 of $\epsilon_K$, $B_d-\overline B_d$ mixing and  $\vert V_{ub}\vert /
\vert V_{cb}\vert$. }
 \end{center}
\end{figure}

In this paper, we have predicted the CKM matrix at the $M_Z$ energy scale
by assuming the F-BS texture for Yukawa matrices at the GUT scale.
In the case of $\tan\beta\simeq 3$, the obtained CKM matrix
are consistent with experiments.
One may worry about our prediction  $V_{cb}\simeq 0.045$ because the
recent experiments favor $V_{cb}= 0.040\pm 0.003$ \cite{Buras}.
There is  a plausible possibly to push down this predicted value.
That is to modify the $SU(5)$ GUT relations by introducing other Higgs fields.
The modification may be guaranteed by the recent lattice calculations of
light quark masses
  \cite{Lattice},
in which  the light quark masses are considerably reduced compared with the
conventional ones.

We emphasize that if the $B$-factory experiments at KEK and SLAC will
restrict the experimentally allowed region of the unitarity triangle
in the first quadrant of the $\rho$-$\eta$ plane,
our proposed simple model can be a candidate for the Yukawa matrices at
the GUT scale.




\end{document}